\documentclass{aa}  
\usepackage{graphicx}
\usepackage{txfonts}
\usepackage{natbib}
\bibpunct{(}{)}{;}{a}{}{,} 
\newcommand{\mincir}{\raise -2.truept\hbox{\rlap{\hbox{$\sim$}}\raise5.truept
\hbox{$<$}\ }}
\newcommand{\magcir}{\raise -2.truept\hbox{\rlap{\hbox{$\sim$}}\raise5.truept
\hbox{$>$}\ }}
\newcommand{\siml}{\raise -2.truept\hbox{\rlap{\hbox{$\sim$}}\raise5.truept
\hbox{$<$}\ }}
\newcommand{\simg}{\raise -2.truept\hbox{\rlap{\hbox{$\sim$}}\raise5.truept
\hbox{$>$}\ }}
\newcommand{\be}{\begin{equation}}
\newcommand{\ee}{\end{equation}}
\newcommand{\ba}{\begin{eqnarray}}
\newcommand{\ea}{\end{eqnarray}}
\newcommand {\kpc} {kpc $\;$}

\newcommand {\h} {Mpc$\;$}

\newcommand {\ks} {km~s$^{-1} \;$}
\newcommand {\kss} {km~s$^{-1}$}
\newcommand {\m} {$M_{\odot} \;$}
\newcommand {\mm} {$M_{\odot}$}

\newcommand {\mqua} {$\times 10^{14}\;M_{\odot} \;$}

\newcommand{\degree}{\ensuremath{\mathrm{^\circ}}}
\newcommand{\arcm}{\ensuremath{\mathrm{^\prime}\;}}
\newcommand{\arcs}{\ensuremath{\arcmm\hskip -0.1em\arcmm \;}}
\newcommand{\arcmm}{\ensuremath{\mathrm{^\prime}}}
\newcommand{\arcss}{\ensuremath{\arcmm\hskip -0.1em\arcmm}}
\newcommand{\dotarcs}{\,\rlap{\hbox{$\mathrm{^\prime\hskip-0.1em^\prime}$}}{\hbox{$.$}}\,}

\newcommand{\dotsec}{\,\rlap{\hbox{$\mathrm{^s}$}}{\hbox{$.$}}\,}

\begin{document}

\title{Optical/X-ray/radio view of Abell 1213: A galaxy cluster with anomalous diffuse radio emission}

\author{
W. Boschin\inst{1,2,3},
M. Girardi\inst{4,6},
S. De Grandi\inst{5},
G. Riva\inst{7,8},
L. Feretti\inst{9},
G. Giovannini\inst{9,10},
F. Govoni\inst{11},
V. Vacca\inst{11}
}


\institute{Fundaci\'on Galileo Galilei - INAF (Telescopio Nazionale Galileo), Rambla Jos\'e Ana Fern\'andez Perez 7, E-38712 Bre\~na Baja (La Palma), Canary Islands, Spain
\and Instituto de Astrof\'{\i}sica de Canarias, C/V\'{\i}a L\'actea s/n, E-38205 La Laguna (Tenerife), Canary Islands, Spain
\and Departamento de Astrof\'{\i}sica, Univ. de La Laguna, Av. del Astrof\'{\i}sico Francisco S\'anchez s/n, E-38205 La Laguna (Tenerife), Canary Islands, Spain
\and Dipartimento di Fisica dell'Universit\`a degli Studi di Trieste - Sezione di Astronomia, via Tiepolo 11, I-34143 Trieste, Italy
\and INAF - Osservatorio Astronomico di Brera, via E. Bianchi 46, I-23807 Merate, Italy
\and INAF - Osservatorio Astronomico di Trieste, via Tiepolo 11, I-34143 Trieste, Italy
\and INAF - Istituto di Astrofisica Spaziale e Fisica Cosmica di Milano, via A. Corti 12, 20133 Milano, Italy
\and Dipartimento di Fisica, Universit\`a degli Studi di Milano, via G. Celoria 16, 20133 Milano, Italy
\and INAF - Istituto di Radioastronomia, Via P. Gobetti 101, 40129 Bologna, Italy
\and Dipartimento di Fisica e Astronomia, Universit\`a di Bologna, Via Gobetti 93/2, 40129 Bologna, Italy
\and INAF - Osservatorio Astronomico di Cagliari, Via della Scienza 5, 09047 Selargius, Italy
}

  \date{Received  / Accepted }

  \abstract {Abell~1213, a low-richness galaxy system, is known to
    host an anomalous radio halo detected in data of the Very Large
    Array (VLA). It is an outlier with regard to the relation between
    the radio halo power and the X-ray luminosity of the parent
    clusters.}
  {Our aim is to analyze the cluster in the optical, X-ray, and radio
    bands to characterize the environment of its diffuse radio
    emission and to shed new light on its nature.}  
  {We used optical data from the Sloan Digital Sky Survey to study the
    internal dynamics of the cluster. We also analyzed archival
    \emph{XMM-Newton} X-ray data to unveil the properties of its hot
    intracluster medium. Finally, we used recent data from the LOw
    Frequency ARray (LOFAR) at 144 MHz, together with VLA data at 1.4
    GHz, to study the spectral behavior of the diffuse radio source.}  
  {Both our optical and X-ray analysis reveal that this low-mass
    cluster exhibits disturbed dynamics. In fact, it is composed of
    several galaxy groups in the peripheral regions and, in
    particular, in the core, where we find evidence of substructures
    oriented in the NE-SW direction, with hints of a merger nearly
    along the line of sight. The analysis of the X-ray emission adds
    further evidence that the cluster is in an unrelaxed dynamical
    state. At radio wavelengths, the LOFAR data show that the diffuse
    emission is $\sim$510 kpc in size. Moreover, there are hints of
    low-surface-brightness emission permeating the cluster center.}
  {The environment of the diffuse radio emission is not what we would
    expect for a classical halo. The spectral index map of the radio
    source is compatible with a relic interpretation, possibly due to
    a merger in the N-S or NE-SW directions, in agreement with the
    substructures detected through the optical analysis. The
    fragmented, diffuse radio emissions at the cluster center could be
    attributed to the surface brightness peaks of a faint central
    radio halo.}


\titlerunning{An optical/X-ray/radio view of Abell 1213} 
\authorrunning{Boschin et al.} 

\maketitle

\section{Introduction}
\label{intro}

Radio halos are diffuse, low-surface-brightness synchrotron sources
($\sim$0.1 $\mu$Jy arcsec$^{-2}$ at 1.4 GHz) hosted in the central
volume of a fraction of massive and unrelaxed galaxy clusters. They
extend up to spatial sizes of 1-2 Mpc, roughly following the
distribution of the intracluster medium (ICM) and have no evident
counterparts in the optical band. Their steep spectrum ($S(\nu)\sim
\nu^{-\alpha}$; with $\alpha>1$) indicates the existence in the ICM of
ultra-relativistic electrons ($\gamma>>1000$) moving in weak ($\sim
\mu$G) magnetic fields (see, e.g., Feretti et al. \cite{fer12} and van
Weeren et al. \cite{wee19} for reviews).

A strong correlation has been found between the total power of radio
halos at 1.4 GHz ($P_{1.4\,{\rm GHz}}$) and the X-ray luminosity
($L_{\rm X}$) of the ICM (Feretti et al. \cite{fer12}, Cassano et
al. \cite{cas13}, Yuan, Han \& Wen \cite{yua15}, Cuciti et
al. \cite{cuc21}), but there are notable outliers. In fact, anomalous
radio halos have been observed with a radio power larger than expected
from the $P_{1.4\,{\rm GHz}}$ - $L_{\rm X,\,0.1-2.4\,kev}$ correlation
shown by the majority of radio halos, with the halo in Abell 523 among
of the most frequently studied of this type (Girardi et
al. \cite{gir16}, Vacca et al. \cite{vac22a} and \cite{vac22b}).

Another controversial case is the diffuse source in the galaxy cluster
\object{Abell 1213} (hereafter A1213). This is a poor (Abell richness
class=1; Abell et al. \cite{abe89}) cluster at $z\sim 0.047$ dominated
in its central region by the FR II radio galaxy 4C29.41 and two more
cluster members identified as radio galaxies.

Very notably, NE of 4C29.41, data of the NRAO VLA Sky Survey (NVSS;
Condon et al. \cite{con98}) and observations with the Very Large Array
(VLA) at 1.4 GHz by Giovannini et al. (\cite{gio09}; hereafter G09)
revealed the presence of a diffuse extended emission that does not
seem due to the discrete radio sources in the cluster central
region. Indeed, G09 classified it as a small-size radio halo,
concluding that the radio morphology and power of this diffuse source
are linked to the physical properties of the cluster as a whole and
not to the activity of cluster radio galaxies. However, if this source
is a radio halo, it would be really peculiar. In fact, it appears well
off-centered with respect to the ICM distribution inferred from
ROSAT/HRI data and overluminous when compared to the ICM X-ray
luminosity (see Fig.~17 of G09). Moreover, the low X-ray luminosity
($L_{\rm X}=0.10\times 10^{44}$ erg s$^{-1}$ in the ROSAT 0.1-2.4 keV
band; Ledlow et al. \cite{led03}) and poor optical richness of A1213
suggest that it is not massive in comparison with typical clusters
known to host a radio halo. An analysis based on redshift data from
the Sloan Digital Sky Survey (SDSS) DR6 by Hern\'andez-Fern\'andez et
al. (\cite{her12}) supports this hypothesis. In fact, they measure a
member galaxies velocity dispersion of $\sim 560$ km s$^{-1}$ (see
their Table~1). This is quite a low value, bearing in mind that
massive clusters with radio halos often exhibit velocity dispersions
$\gtrsim$1000 km s$^{-1}$.

In this intriguing context, we performed an exhaustive study of A1213
in the optical, X-ray, and radio bands. In particular, we used SDSS
photometric and spectroscopic information to extend the analysis of
Hern\'andez-Fern\'andez et al. (\cite{her12}) by searching for optical
substructures, which are an indicator of the dynamical state of the
cluster (e.g., Girardi \& Biviano \cite{gir02}). Then, we complemented
the optical analysis with public X-ray data from \emph{XMM-Newton} to
draw a multiband picture of A1213 and use this information to study
the environment of its extended radio emission. Finally, we used
archival radio data from the LOw Frequency ARray (LOFAR) and the VLA
data by G09 to derive the spectral properties of the diffuse radio
source.

The paper is organized as follows. We describe the analyzed optical
data and select the members of the cluster in Sect.~2. We estimate the
global properties and analyze the optical substructures in Sects.~3
and ~4. In Sect.~5, we discuss the large-scale structure around the
cluster, while in Sect.~6, we present the analysis and results of the
\emph{XMM-Newton} X-ray data. In Sect.~7, we show the radio data from
LOFAR. Section~8 is devoted to the summary and discussion of our
results. In this work, we use $H_0=70$ km s$^{-1}$ Mpc$^{-1}$ in a
flat cosmology with $\Omega_{\rm m}=0.3$ and
$\Omega_{\Lambda}=0.7$. In the assumed cosmology, 1\arcm corresponds
to $\sim 55.2$ \kpc at the cluster redshift. We recall that the
velocities we derive for the galaxies are line-of-sight velocities
determined from the redshift, $V=cz$. Unless otherwise stated, we
report errors with a confidence level (c.l.) of 68\%.

\section{Redshift data and spectroscopic cluster members}
\label{memb}

From SDSS DR13, we extracted 402 galaxies with a redshift $z\le0.2$
within a radius $R=60$\arcmm, corresponding to 3.31 Mpc in the cluster
rest frame, from the position
R.A.=$11^{\mathrm{h}}16^{\mathrm{m}}40\dotsec00$, Dec.=$+29\degree
16\arcmm 00\dotarcs0$ (J2000).

\begin{figure}[!ht]
\centering
\resizebox{\hsize}{!}{\includegraphics{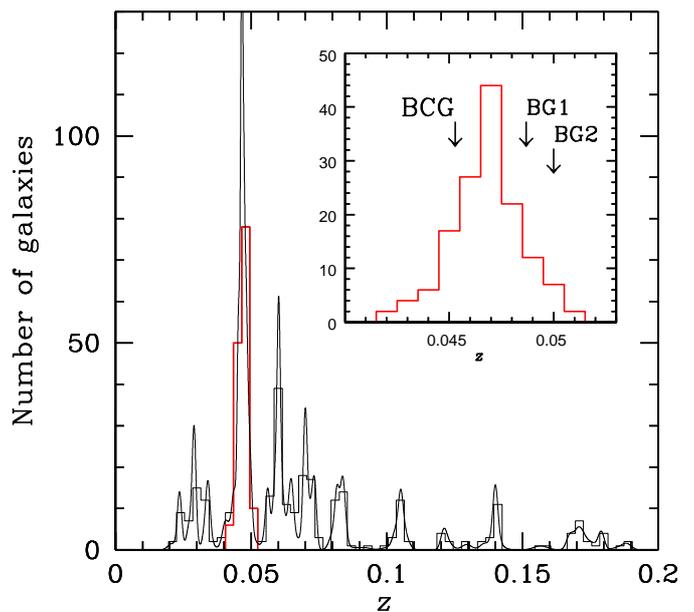}}
\caption
{Distribution of redshifts of galaxies with $z\leq 0.2$. The histogram
  refers to all galaxies with spectroscopic redshifts in the region of
  A1213. The histogram with the thick red line highlights the 144
  galaxies that are assigned to A1213 using the 1D-DEDICA
  reconstruction method. The inset shows the final 143 member
  galaxies, with the redshifts of the BCG and the pair of bright close
  galaxies BG1 and BG2 shown.}
\label{fighisto}
\end{figure}

To select cluster members among the 402 galaxies in the spectroscopic
catalog, we used the two-step method known as Peak+Gap (P+G), which
was previously applied by Girardi et al. (\cite{gir15}). The first
step is the application of the adaptive-kernel DEDICA method (Pisani
\cite{pis93} and Pisani \cite{pis96}; see also Girardi et
al. \cite{gir96}). Using this method, A1213 is identified as a peak at
$z\sim0.04658$ consisting of 144 galaxies in the range $0.041989 \leq
{\rm z} \leq 0.052006$, that is, $12588 \leq{\rm V} \leq 15591$ \ks
for the line--of--sight velocity $V=cz$ (see Fig.~\ref{fighisto}). Of
the non-member galaxies, 53 are foreground and 205 are background
galaxies.
        
\begin{figure}[!ht]
\centering
\resizebox{\hsize}{!}{\includegraphics{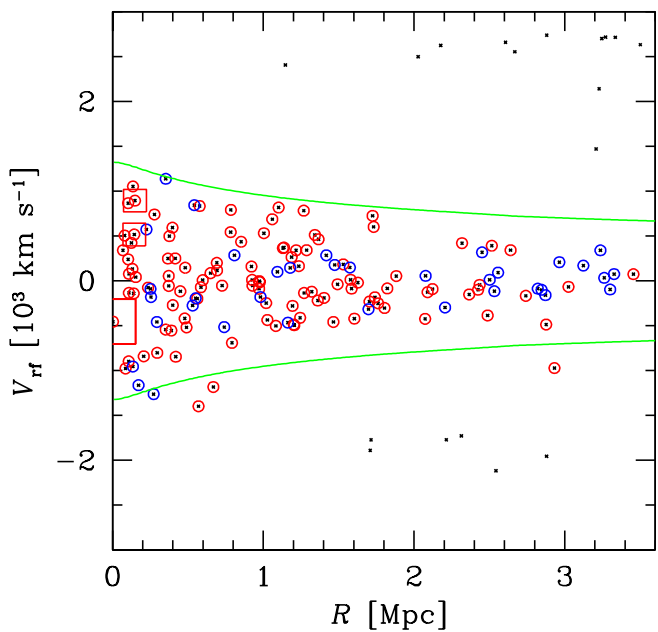}}
\caption
    {Projected phase space of the galaxies of A1213, where the rest
      frame velocity $V_{\rm rf}=(V-\left<V\right>)/(1+z)$ versus the
      projected cluster-centric distance $R$ is plotted. Only galaxies
      with redshifts in the range $\pm 3000$ \ks are shown (black
      dots). The red and blue circles show the 143 members of the
      cluster (red and blue galaxies, as defined in the text). The
      giant and large red squares refer to the BCG and the pair of
      bright galaxies BG1 and BG2, respectively. The green curves
      contain the region where $|V_{\rm rf}|$ is smaller than the
      escape velocity (see text).}
\label{figvd}
\end{figure}

In a second step, we combine the space and velocity information by
using the ``shifting gapper'' method (Fadda et al. \cite{fad96},
Girardi et al. \cite{gir96}). Of the galaxies that lie within an
annulus around the center of the cluster, this procedure rejects those
that are too far away in terms of velocity from the main body (i.e.,
farther away than a fixed velocity gap). The position of the annulus
is shifted with increasing distance from the center of the
cluster. The procedure is repeated until the number of cluster members
converges to a stable value. Following Fadda et al. (\cite{fad96}), we
used a gap of $1000$ \ks in the cluster rest frame and an annulus size
of 0.6 \h or more to include at least 15 galaxies. In determining the
cluster center, we considered the position in right ascension (R.A.)
and declination (Dec.) of the brightest cluster galaxy in our sample
(BCG) [R.A.=$11^{\mathrm{h}}16^{\mathrm{m}}22\dotsec70$,
  Dec.=$+29\degree 15\arcmm 08\dotarcs3$ (J2000)]. With this
procedure, one galaxy is discarded and we confirm 143 cluster
members. The distribution of member galaxies is shown in the redshift
space and in the projected phase space in Fig.~\ref{fighisto} and
Fig.~\ref{figvd}, respectively. To highlight the region of cluster
members, we also plot in Fig.~\ref{figvd} the escape velocity curves
obtained with the mass estimate calculated below (see
Sect.~\ref{prop}), assuming a Navarro, Frenk \& White (\cite{nav97})
mass-density profile and adopting the prescription of den Hartog \&
Katgert (\cite{den96}).

\begin{figure*}[!ht]
\centering 
\includegraphics[width=18cm]{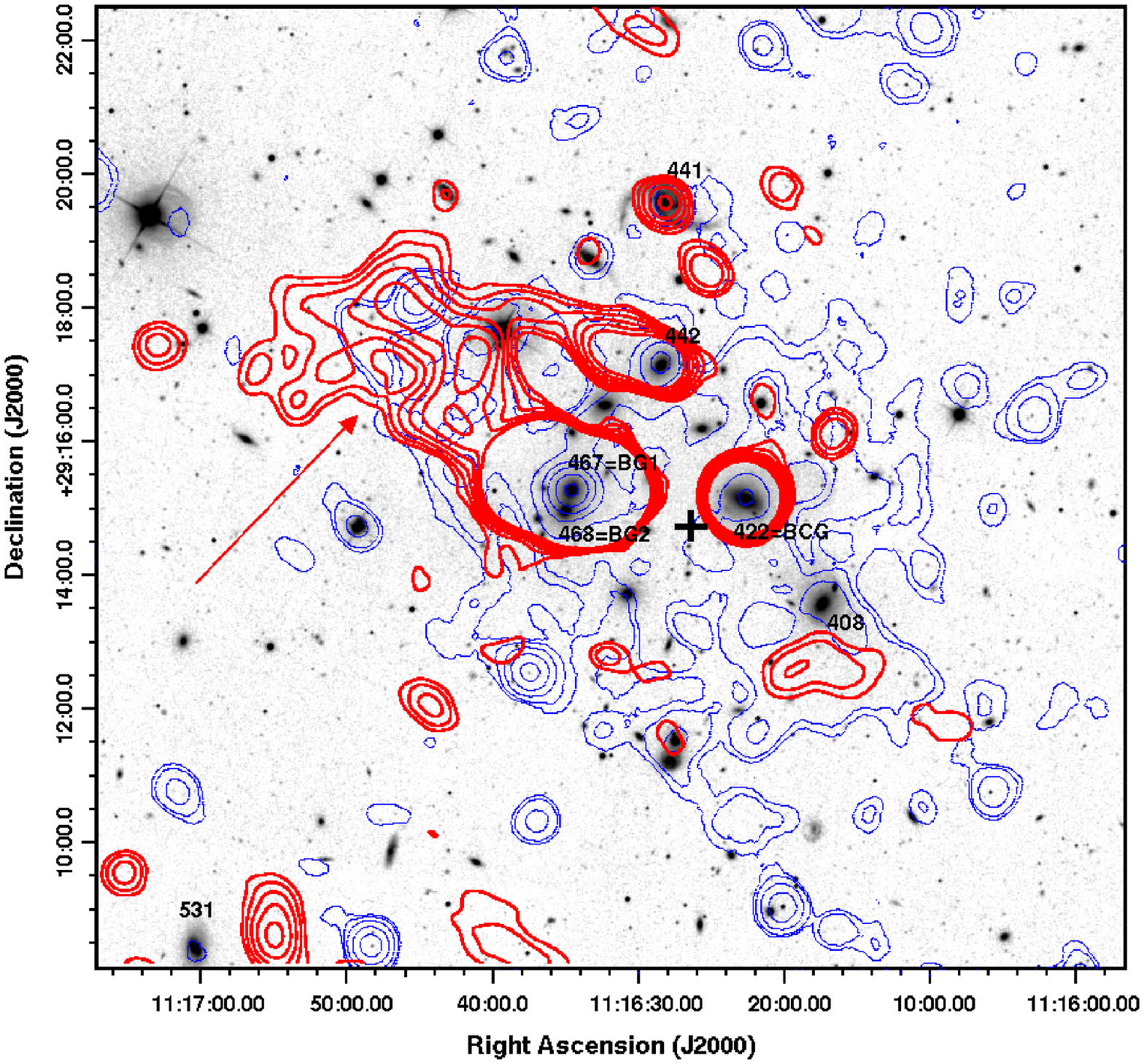}
\caption{SDSS $g$-band image of the galaxy cluster A1213 with,
  superimposed, the contour levels of the VLA 1.4 GHz radio image by
  G09 (red thick contours, HPBW=$35\arcss\times35\arcss$, first
  contour level at 1 mJy/beam, the others spaced by a factor
  $\sqrt{2}$). Thin blue contours show the (smoothed) X-ray emission
  of the cluster as derived from the \emph{XMM-Newton} archival image
  ID~0550270101 (photons in the energy range 0.7--1.2 keV). Labels
  denote galaxies discussed in the text. The arrow highlights the
  diffuse radio emission classified by G09 as a small-size radio
  halo. The black cross shows the position of the centroid of the
  X-ray emission (see Sect.~\ref{xray}).}
\label{figottico}
\end{figure*}

Table~\ref{catalogA1213} lists the velocity catalog for the prominent
member galaxies. Instead, Fig.~\ref{figottico} shows the central
region of the cluster with, superimposed, the VLA radio (from G09) and
\emph{XMM-Newton} X-ray contours obtained from our analysis (see
Sect.~\ref{xray}).

The cluster galaxy population is dominated by the brightest cluster
galaxy (BCG) ID~422, which is $\sim70$\arcs far from the centroid of
the X-ray emission (see Sect.~\ref{xray}). It is also a bright radio
source. However, our catalog lists several galaxies with magnitude
$r\leq r_{\rm BCG}+1$. Indeed, A1213 is a typical ``core'' cluster, as
classified by Rood \& Sastry (\cite{roo71}).

In the central region, the most interesting galaxies are the couple
IDs~467+468 (see Fig.~\ref{figottico}), very close in the sky and
separated by only 400 \ks in radial velocity. Notably, they show an
excess of intracluster light likely due to interaction, thus we treat
them as the bright couple BG1+BG2 (see Table~\ref{catalogA1213}). If
we sum their fluxes we obtain a magnitude r$\sim$13.75. They have an
higher velocity with respect to the average and act as a counterpoint
to the BCG (see Fig.~1). We note that ID~467=BG1 is the optical
counterpart of the FRII radio galaxy 4C29.41 (see G09 and
Sect.~\ref{intro}).

In Fig.~\ref{figottico}, we highlight more notable galaxies. At $\sim
4.6$\arcm NNE of the BCG, ID~441 is a star-forming spiral
galaxy. $\sim 2$\arcm S of ID~441, ID~442 is an head-tail radio galaxy
oriented toward NE. More bright galaxies in Fig.~\ref{figottico} are
IDs.~408 and 531.

Far from the cluster center, at $\sim 23$\arcm NW of the BCG (not
shown in Fig.~\ref{figottico}), ID~274 is a prominent cluster member
with a close bright companion. We do not have a spectroscopic redshift
for the latter, but the photometric redshift estimate from the SDSS
($z_{\rm phot}\sim 0.050\pm 0.009$) suggests it is also a likely
cluster member.

\section{Global properties and galaxy population}
\label{prop}

The analysis of the velocity distribution of the 143 cluster members
was performed using the biweight estimators for location and scale
included in ROSTAT (statistical routines of Beers et
al. \cite{bee90}). Our measurement of the mean redshift of the cluster
is $\left<z\right>=0.0469\pm0.0001$ (i.e., $\left<V\right>=14052\pm39$
\kss). We estimate the velocity dispersion, $\sigma_{\rm V}$, by
applying the cosmological correction and the standard correction for
velocity errors (Danese et al. \cite{dan80}). We obtained a value of
$\sigma_{\rm V}=463_{-31}^{+41}$ \kss, where the errors are estimated
by a bootstrap technique.

To derive the mass\footnote{We refer to $R_{\Delta}$ as the radius of
  a sphere in which the average mass density is $\Delta$ times the
  critical density $\rho_{\rm c}$ at the redshift of the galaxy
  system; $M_{\Delta}=(4/3)\pi \Delta \rho_{\rm c}R_{\Delta}^3$ is the
  mass contained in $R_{\Delta}$.} $M_{200}$, we used the theoretical
relation between the mass $M_{200}$ and the velocity dispersion in
clusters presented and verified with simulated clusters by Munari et
al. (\cite{mun13}; their Eq.~1 and Fig.~1). We took a recursive
approach to this step. To obtain a first estimate of the radius
$R_{200}$ and $M_{200}$, we applied the relation of Munari et
al. (\cite{mun13}) to the global value of $\sigma_{\rm V}$ obtained
above. We considered the galaxies within this first estimate of
$R_{200}$ to recalculate the velocity dispersion. The procedure is
repeated until a stable result is obtained. We estimated $\sigma_{\rm
  V,200}=573_{-38}^{+52}$ for 81 galaxies within
$R_{200}=1.20_{-0.08}^{+0.11}$ Mpc, in good agreement with the
estimate by Hern\'andez-Fern\'andez et al. (\cite{her12}). The mass is
$M_{200}=2.0_{-0.6}^{+0.4}$ \mqua. The uncertainty for $R_{200}$ is
calculated using the error propagation for $\sigma_{\rm V}$
($R_{200}\propto \sigma_{\rm V}$), while the uncertainty for $M_{200}$
is computed considering that $M_{200}\propto \sigma_{\rm V}^3$ and
adding an uncertainty of $10\%$ due to the scatter around the relation
of Munari et al. (\cite{mun13}). The global properties of the cluster
are shown in Table~\ref{tabv}.

\begin{figure}
\centering
\resizebox{\hsize}{!}{\includegraphics{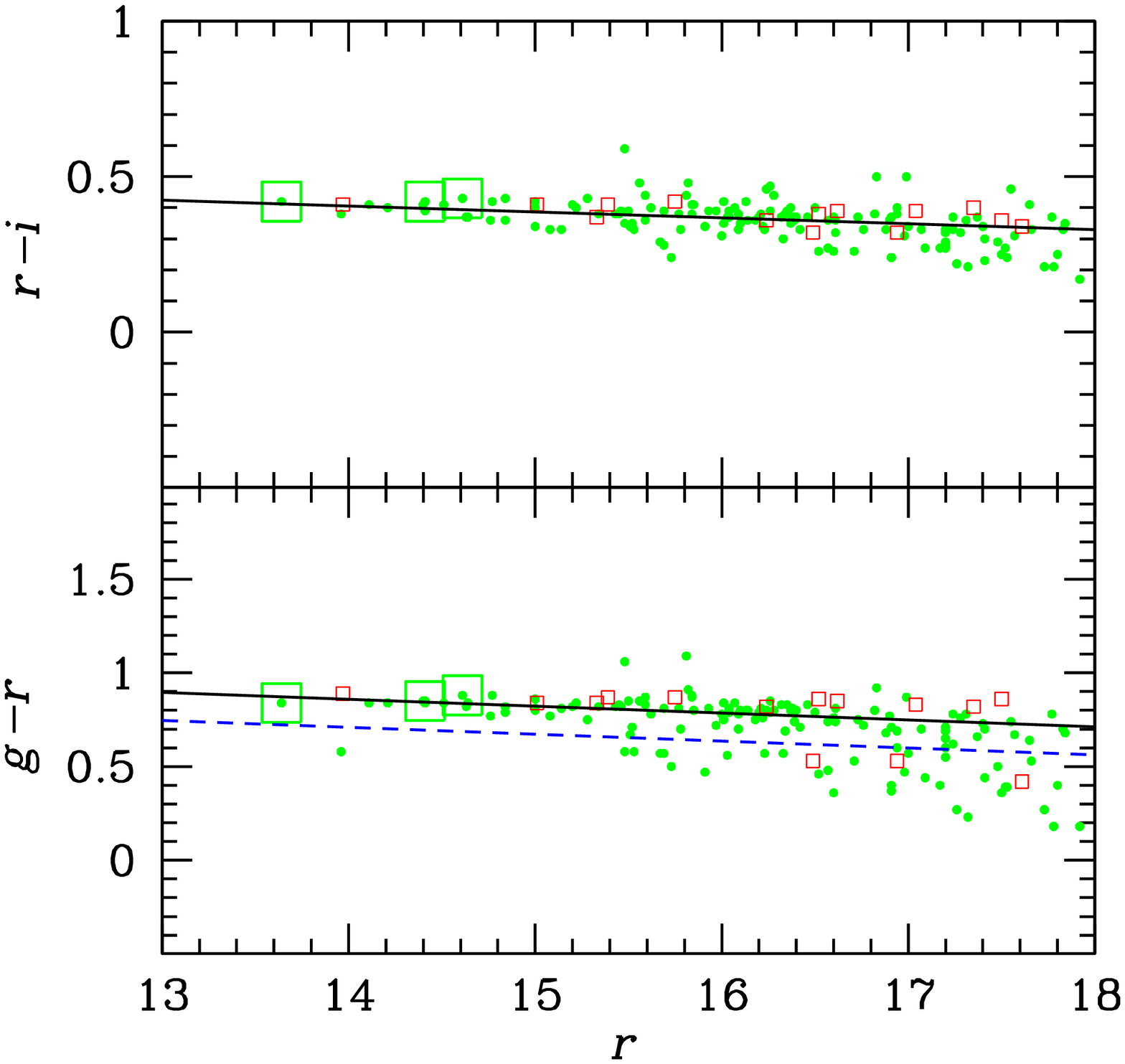}}
\caption
    {PanSTARRS magnitude diagrams: $r$$-$$i$ vs. $r$ ({\em upper
        panel}) and $g$$-$$r$ vs. $r$ ({\em lower panel}) of member
      galaxies of A1213 (green points) and the background eastern
      group (red squares, see Sect.~\ref{back}).  Large green squares
      show the BCG and the pair of bright galaxies: BG1 and BG2.
      Black lines show the color-magnitude relations obtained for the
      A1213 member galaxies.  The blue dashed line in the lower panel
      is used to separate red galaxies from blue ones.}
\label{figcm}
\end{figure}

Figure~\ref{figcm} shows the distribution of member galaxies in the
($r$$-$$i$ vs. $r$) and ($g$$-$$r$ vs. $r$) color-magnitude
diagrams. The ($g$$-$$r$ vs. $r$) color-magnitude relation indicating
the location of early-type galaxies is seen down to faint magnitudes
of $r\sim$17.5 mag, about 2.5 mag fainter than the $m_{r}^*$
value. Following Boschin et al. (\cite{bos12}) we use a $2\sigma$
rejection procedure to obtain $r$$-$$i$=0.671$-$0.019$\times$ $r$ and
$g$$-$$r$=1.378$-$0.037$\times$ $r$. We used the ($g$$-$$r$ vs. $r$)
relation to classify red and blue galaxies (107 and 36, respectively).
Blue galaxies are defined as galaxies that are 0.15 mag bluer than the
color expected for their magnitude in the red sequence (i.e., the
dashed blue line in Fig.~\ref{figcm}, lower panel; e.g., Boschin et
al. \cite{bos20}). Other galaxies are defined as red galaxies.

\section{Analysis of optical substructure}
\label{sub}

We analyzed the presence of substructures using the velocity
distribution of galaxies, their projected positions on the sky, and
the combination of these two pieces of information (1D, 2D, and 3D
tests, respectively).

Using a set of indicators such as kurtosis, skewness, tail index, and
asymmetry index (Bird \& Beers \cite{bir93}), the analysis of the
velocity distribution shows no evidence of any possible deviations
from the Gaussian distribution. The BCG has a lower velocity than the
mean velocity of the cluster, and Fig.~\ref{fighisto} (inset) shows
how much this velocity is peculiar within the velocity
distribution. According to the Indicator test by Gebhardt \& Beers
(\cite{geb91}), the BCG velocity is peculiar at the $>99\%$ c.l.; the
same result is obtained when considering the sample of galaxies within
$R_{200}$. For the BCG, we calculate $|V_{\rm rf}/\sigma_{\rm
  V,200}|=0.80$, placing A1213 in the upper range of $|V_{\rm
  rf}/\sigma_{\rm V}|$ distribution, since only about 10\% of the
clusters have such a high value (see Fig.~8 of Lauer et
al. \citealt{lau14}).

\begin{figure}
\resizebox{\hsize}{!}{\includegraphics{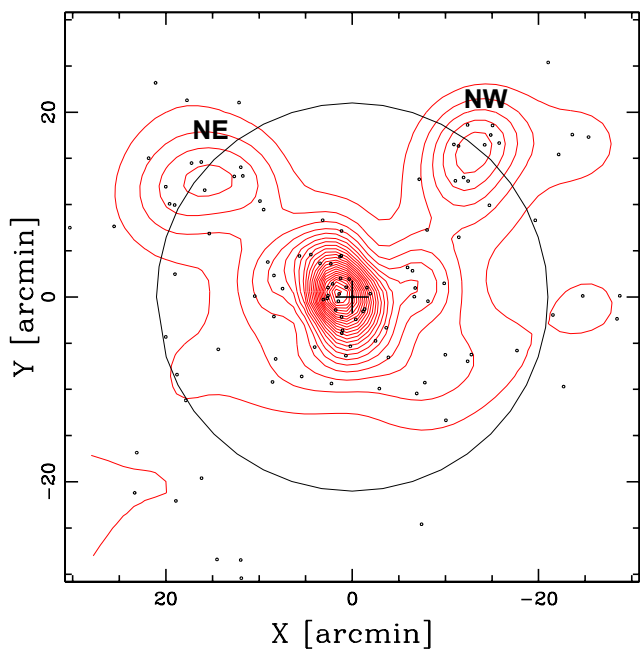}}
\caption{Spatial distribution in the plane of the sky of the cluster
  members (black dots) with the corresponding isodensity contour map
  obtained with the 2D-DEDICA method (red contours). The black cross
  shows the position of the BCG, which is considered to be the center
  of A1213. The diagram is centered on the center of the cluster. A
  circle with a radius of 21\arcmm, i.e., approximately $R_{200}$, is
  drawn. The labels refer to the two peripheral 2D-DEDICA peaks listed
  in Table~\ref{tabdedica2dz}.}
\label{figk2z}
\end{figure}

Then, we analyzed the spatial 2D distribution of the 143 member
galaxies (see Fig.~\ref{figk2z}). First, we computed the ellipticity
($\epsilon$) and position angle of the major axis ($PA$, measured
counterclockwise from north) using the method based on moments of
inertia (Carter \& Metcalfe \cite{car80}; see also Plionis \&
Basilakos \cite{pli02}, where weight equals one). We obtained
$\epsilon=0.24_{-0.10}^{+0.06}$ and ${\rm PA}=119_{-10}^{+11}$. The
low value of the ellipticity is due to the complex structure of the
cluster rather than a round, homogeneous distribution of galaxies, as
shown below.

We also used the 2D-DEDICA method (Pisani \cite{pis96}, see also
Girardi et al. \cite{gir96}). Our results are shown in
Fig.~\ref{figk2z} and Table~\ref{tabdedica2dz}.  For each detected
galaxy clump with a c.l. greater than 99\% and a relative density
(with respect to the densest peak) $\rho\gtrsim 0.2$,
Table~\ref{tabdedica2dz} lists the number of member galaxies, the
position of the 2D density peak, and $\rho$. Figure~\ref{figk2z} shows
the main peak in the cluster core (2D-CORE in
Table~\ref{tabdedica2dz}) and two peaks in the NW and NE (2D--NW and
2D--NE) external regions of the cluster (at $R\sim R_{200}$). The BCG
is contained in the cluster core, while the NW peak is dominated by
the luminous galaxy ID~274 and a close bright likely cluster member
(see Sect.~\ref{memb}). The NE peak follows the direction of the
main elongation of the X-ray isophotes (see Sect.~\ref{xray}). For all
groups, Table~\ref{tabdedica2dz} also lists the mean velocities, which
do not significantly differ taking into account the relative
uncertainties. We note that the velocity of the BCG is peculiar with
respect to the velocity distribution in the 2D-CORE clump at the 99\%
c.l.

\begin{figure}
\centering 
\includegraphics[width=8cm]{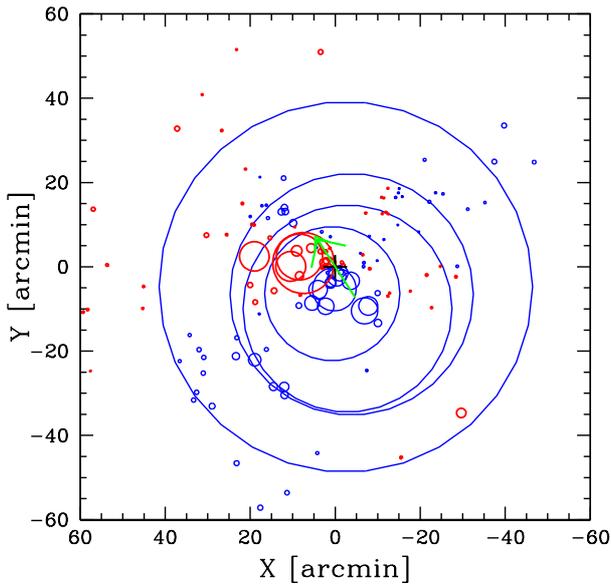}
\caption
{Dressler \& Schectman bubble plot for the DSV test. Spatial
  distribution of the 143 cluster members, each indicated by a symbol:
  the larger the symbol, the larger the deviation $|\delta_{i,{\rm
      V}}|$ of the local mean-velocity from the global mean-velocity.
  The blue thin and red thick circles indicate where the local mean
  velocity is smaller or larger than the global mean velocity. Here
  the bubble size is less enhanced than the standard size for better
  readability (size equal to exp$(w/3)$, with $w=|\delta_{i,{\rm
      V}}|$). The green arrow indicates the position angle of the
  velocity gradient that is calculated within 0.5$R_{200}$. The
  diagram is centered on the BCG.}
\label{figds10v}
\end{figure}

In the full 3D analysis, we looked for a correlation between velocity
and position data. The eventual presence of a velocity gradient is
quantified by a multiple linear regression fit to the observed
velocities with respect to galaxy positions in the plane of the sky
(see also den Hartog \cite{den96}). To assess the significance of this
velocity gradient, we performed 1000 Monte Carlo simulations of galaxy
clusters by randomly shuffling the velocities of the galaxies and
determining, for each simulation, the coefficient of multiple
determination ($RC^2$, NAG Fortran Workstation Handbook
\cite{nag86}). The significance of the velocity gradient is the
fraction of cases in which the $RC^2$ of the simulated data is smaller
than the observed $RC^2$. In A1213, the velocity gradient is not
significant in the whole sample as well as within the $R_{200}$
region. Within the $0.5R_{200}$ region, the velocity gradient is
significant at the $98\%$ confidence level.  In this case, the
position angle on the celestial sphere is $PA=33_{-16}^{+12}$ degrees,
which means that the high-velocity galaxies are located in the NE
region of the cluster (see Fig.~\ref{figds10v}).

We then used our modified version of the $\Delta$-test of Dressler \&
Shectman (\cite{dre88}), which considers only the indicator of local
mean velocity (hereafter DSV test, Girardi et al. \cite{gir16}). This
indicator is $\delta_{i,{\rm V}}=[(N_{\rm nn}+1)^{1/2}/\sigma_{\rm
    V}]\times (V_{\rm loc} -\left<V\right>)$, where the local mean
velocity $V_{\rm loc}$ is calculated using the $i$-th galaxy and its
$N_{\rm{nn}}=10$ neighbors. For a cluster, the cumulative deviation is
given by the value of $\Delta$, which is the sum of the
$|\delta_{i,{\rm V}}|$ values of the individual $N$ galaxies. As in
the calculation of the velocity gradient, the significance of the
$\Delta$ (i.e., the presence of substructure) is based on 1000 Monte
Carlo simulated clusters. In A1213, the significance of the
substructure is 96.3\%. In Fig.~\ref{figds10v}, we show the Dressler \&
Schectman bubble-plot resulting from the indicator of the DSV test for
the whole sample. Larger circles indicate galaxies where the local
mean velocity deviates more from the global velocity. The visual
inspection of the bubble plot suggests the presence of poorly
populated substructures in the core, in particular, a low-velocity
substructure in the SW (large blue circles) and possibly a
high-velocity substructure in the NE (large red circles). The relative
position of these substructures is consistent with our estimate of the
$PA$ of the velocity gradient in the central cluster region.

\begin{figure}
\centering 
\resizebox{\hsize}{!}{\includegraphics{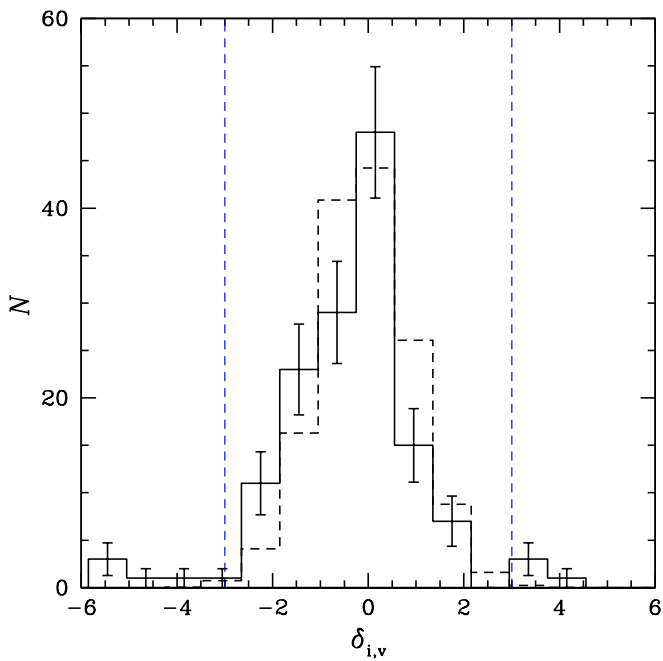}}
\caption
    {Distribution of $\delta_{i,{\rm V}}$ values of the deviation of
      the local mean velocity from the global velocity (according to
      the DSV test, see text).  The histogram with the solid line
      shows the observed galaxies. The histogram with the dashed line
      shows the galaxies of the simulated clusters, normalized to the
      number of observed galaxies. The blue vertical lines indicate
      the $|\delta_{i,{\rm V}}|>3$ regions where we expect to find
      substructure members.}
\label{figdeltai}
\end{figure}

As an attempt to detect substructure members, we resorted to the
technique developed by Biviano et al. (\cite{biv02}). We compared the
distribution of $\delta_{i,{\rm V}}$ values of the real galaxies with
the distribution of $\delta_{i,{\rm V}}$ values of the galaxies of all
the 1000 Monte Carlo simulated clusters (Fig.~\ref{figdeltai}). The
distribution of values of real galaxies shows a tail at low
$\delta_{i,{\rm V}}$ values and possibly a tail at high
$\delta_{i,{\rm V}}$ values. Looking at galaxies with $|\delta_{i,{\rm
    V}}|\gtrsim 3$, we found five galaxies in the low tail and four
galaxies in the high tail, with none of them being a luminous galaxy
or a galaxy that belongs to the prominent galaxies
(Sect.~\ref{memb}).

\begin{figure}
\centering 
\includegraphics[width=8cm]{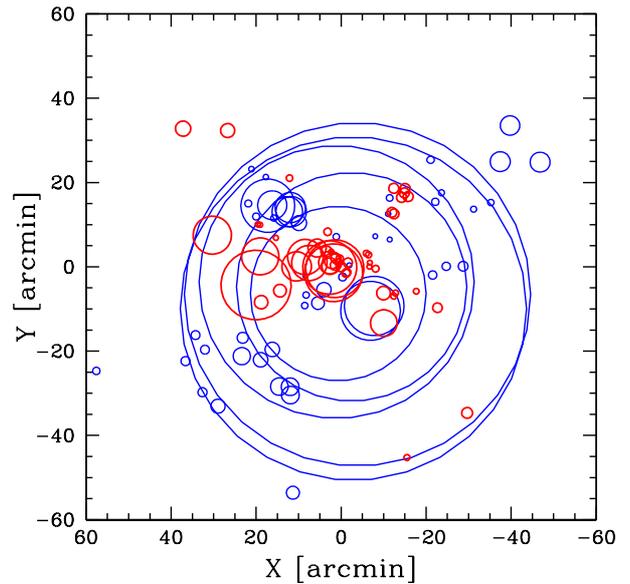}
\caption
    {Dressler \& Schectman bubble plot for the DSV test (see also
      Fig.~\ref{figds10v}) applied to the red galaxy population only
      and looking for small substructures ($N_{\rm{nn}}=5$). Here, the
      bubble size follows the traditional exponential scale (size
      equal to exp$(w)$, with $w=|\delta_{i,{\rm V}}|$).}
\label{figds5vred}
\end{figure}

Since the substructures seem to consist of few galaxies, we repeated
the DSV test with $N_{\rm{nn}}=5$ neighbors. The significance of the
substructure is 96.6\% for the whole sample. If we consider only red
passive galaxies, which usually trace the most important structures
and merger remnants, the significance increases to 98.3\%. In
Fig.~\ref{figds5vred}, we show the Dressler \& Schectman bubble-plot
obtained for the sample of red galaxies. We confirm the presence of a
low-velocity poorly populated substructure in the core.

\begin{figure}
\centering
\resizebox{\hsize}{!}{\includegraphics{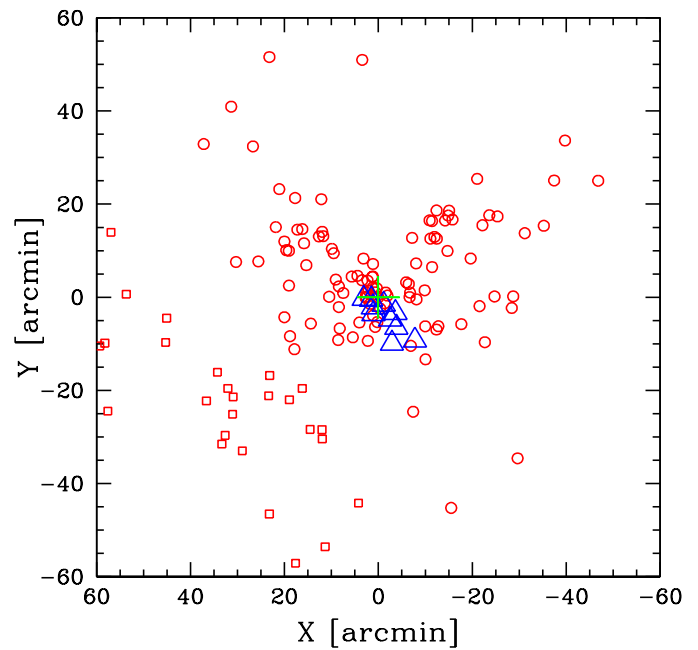}}
\caption{Spatial distribution of the 143 cluster members in the sky,
  with the groups discovered with the 3D-DEDICA method marked by
  different symbols. Red circles and squares show the galaxies of
  3D--NW and 3D--SE peaks, respectively. Blue triangles indicate the
  galaxies of the 3D-CORE peak, which are characterized by low
  velocities. The green central cross indicates the position of the
  BCG.}
\label{figded3d}
\end{figure}

Finally, we used the 3D-DEDICA method (Pisani \cite{pis96}; Bardelli
et al. \cite{bar98}), which splits A1213 into more than ten groups.
Since it is well known that the multidimensional application of the
DEDICA algorithm could lead to spurious substructure (Bardelli et
al. \cite{bar98}), we also applied the alternative version of Balestra
et al. (\cite{bal16}), based on the rule of thumb for the kernel size
given by Silverman (\cite{sil86}). The goal here is to identify the
most important substructures at the expense of losing some smaller
substructures. Table~\ref{tabdedica3d} lists the properties of all the
galaxy peaks detected with a probability higher than the $>99\%$
c.l. and a relative density $\rho\ge 0.2$. Figure~\ref{figded3d} shows
the positions of the galaxies associated with the detected peaks. The
galaxies assigned to the 3D--NW and 3D--SE peaks form the main system
and have similar mean velocities, i.e., they are probably split
because of their different 2D positions. The 3D-CORE peak is a further
confirmation of the existence of a group with low-velocity galaxies
and projected onto the core.

\section{Large scale structure around A1213}
\label{back}

\begin{figure}[!hb]
\centering
\resizebox{\hsize}{!}{\includegraphics{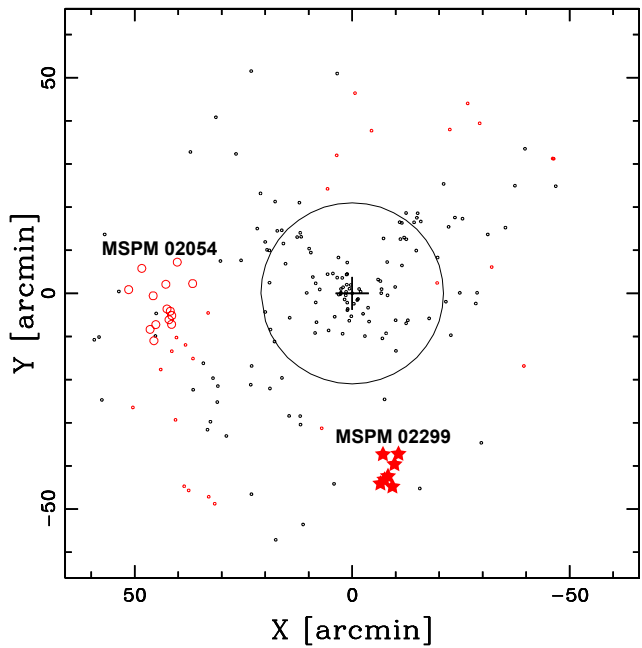}}
\caption
    {Spatial distribution of the 47 galaxies populating the redshift
      peak at $z\sim 0.06$ (red symbols). The galaxies highlighted by
      red circles are assigned to the eastern clump by 2D-DEDICA,
      while red stars put in evidence galaxies belonging to the group
      MSPM 02299 (see text). Black points are the 143 members of
      A1213, whose BCG is indicated by a black cross. A circle with a
      radius of 21$\arcm$ ($\sim R_{200}$) is also drawn.}
\label{figpiccop6}
\end{figure}

From our analysis of the redshift distribution (see
Fig.~\ref{fighisto}), the second peak in relative density is at
$z\sim0.06$ and it is populated by 47 galaxies. Using the 2D-DEDICA
method, we find that part of its galaxy population (14 galaxies)
clumps together at ESE around the coordinates
R.A.=$11^{\mathrm{h}}19^{\mathrm{m}}35\dotsec00$, Dec.=$+29\degree
09\arcmm 48\dotarcs7$ (J2000; see Fig.~\ref{figpiccop6}). For this
group of 14 galaxies, we estimate a velocity dispersion of
$\sigma_{\rm V}\sim 300$ \ks and $\left<z\right>=0.0601$.

The redshift difference between A1213 and this group may be
interpreted as a luminosity distance of 61 Mpc or that the group is
moving at $V_{\rm rf}\sim 4000$ \ks with respect to the A1213 cluster
frame. This speed is of the order of the core-core falling speed in
massive clusters. Since A1213 is a poor cluster, we think they are not
gravitationally interacting. Moreover, the group is located at 42\arcm
from the center of A1213, i.e. at a projected distance $>$2 Mpc. Also,
we note that the color $g-r$ of the group galaxies is slightly redder
than the color of red-sequence galaxies of A1213 (see
Fig.~\ref{figcm}), which is another hint that the group is in the
background with respect to A1213.

According to NED\footnote{The NASA/IPAC Extragalactic Database (NED)
  is operated by the Jet Propulsion Laboratory, California Institute
  of Technology, under contract with the National Aeronautics and
  Space Administration.}, the center of the background group coincides
with WHL J111934.5+291100 (also MSPM 02054), at $z\sim 0.059$. We note
that at $\sim$40\arcm SSW of A1213, an handful of galaxies (see
Fig.~\ref{figpiccop6}) defines the galaxy group MSPM 02299 ($z\sim
0.061$, from NED). At 1.4\degree$\,$SW of A1213 (not shown in
Fig.\ref{figpiccop6}), Abell 1185 is the most outstanding structure in
the surroundings. According to Einasto et al. (\cite{ein01}), A1213
and Abell 1185 (at $z\sim$0.033) do not belong to the same
supercluster.

\section{Analysis of the X-ray data}
\label{xray}

A1213 was observed by the European Photon Imaging Camera (EPIC, Turner
et al. \cite{tur01}) instrument on board the \emph{XMM-Newton}
satellite. EPIC is formed of three detectors MOS1, MOS2, and pn that
simultaneously observe the same target. The archival observation was
pointed on the radio galaxy 4C29.41, whose optical counterpart is
ID~467=BG1 (obs. ID 0550270101). We reprocessed the dataset using the
Extended-Science Analysis System (ESAS, Snowden et al. \cite{sno08})
embedded in SAS version 16.1 following the analysis described in
detail in Ghirardini et al. (\cite{ghi19}). After the soft protons
cleaning procedure (with the {\it mos-filter} and {\it pn-filter} ESAS
tasks; Snowden et al. \cite{sno08}), the total available clean
exposure time is 16.5 ks for MOS1, 14.8 ks for MOS2, and 4.3 ks for
pn.

\begin{figure}
\centering
\resizebox{\hsize}{!}{\includegraphics{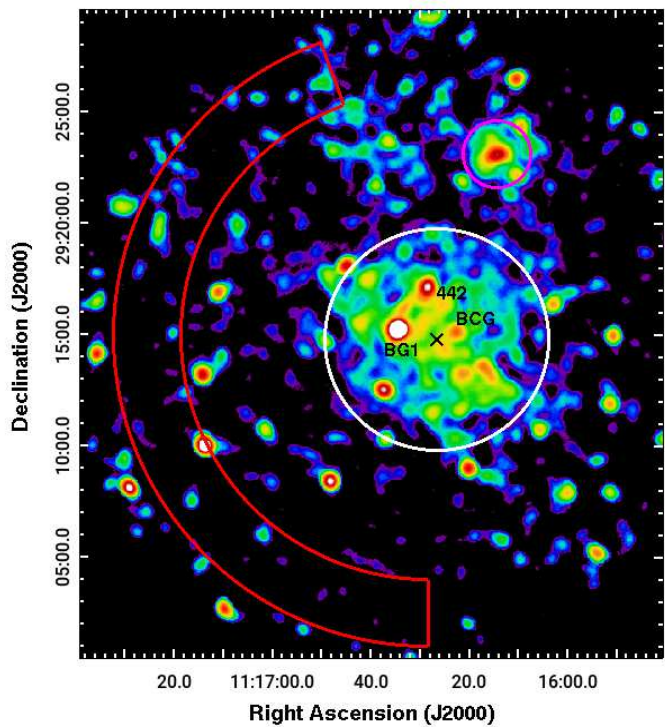}}
\caption{Smoothed, vignetting-corrected \emph{XMM-Newton}/EPIC
  count-rate image of A1213 in the 0.7-1.2 keV band. The circles show
  the regions used to estimate the global X-ray properties of A1213
  (white) and of the extended source located NNW of the cluster
  (magenta, see text). Instead, the red region was used to estimate
  the local background components. The black cross marks the centroid
  of the X-ray emission. Black labels highlight galaxies that are also
  X-ray point sources.}
\label{figX1}
\end{figure}

We extracted, for each EPIC detector, the photon-count images in the
0.7$-$1.2 keV band, which is the energy band that maximizes the
source-to-background ratio in galaxy clusters (e.g., Ettori et
al. \cite{ett10}), and co-added them to obtain a total EPIC
image. With the ESAS tool {\it eexpmap}, we produced the EPIC exposure
map folding the vignetting effect. We used the ESAS collection of
closed-filter observations to produce a map of the non-X-ray
background (NXB), which we rescaled to our observations by comparing
the count rates in the unexposed corners of the field of
view. Figure~\ref{figX1} shows the resulting vignetting-corrected,
NXB-subtracted, count rate image of A1213. A Gaussian filter smooths
the image for visual purposes only.

\subsection{X-ray properties of A1213}
\label{xray1}
We performed a spectral analysis of A1213 following the analysis
described in Ghirardini et al. (\cite{ghi19}). For each region, we
extracted spectra and response files using the ESAS tasks {\it
  mos-spectra} and {\it pn-spectra}. From the spectra we filtered out
the point sources detected in the field of view by running the SAS
wavelet detection tool {\it ewavdetect}.

Since the surface brightness of cluster A1213, apart from some bright
point sources, appears to be very shallow, following Leccardi \&
Molendi (\cite{lec08}) and Ghirardini et al. (\cite{ghi19}) we
preferred to model the background instead of subtracting it. This
approach models the background with different spectral components: (a)
the NXB component and (b) the sky-background component, estimated from
the region shown in red in Fig.~\ref{figX1} (see Ghirardini et
al. \cite{ghi19} for a detailed explanation of both components).

To estimate the global properties of A1213 we extracted a spectrum
from a circle with 5 arcmin radius (shown in white in
Fig.~\ref{figX1}) centered on the centroid of the X-ray emission
(R.A.=$11^{\mathrm{h}}16^{\mathrm{m}}26\dotsec40$, Dec.=$+29\degree
14\arcmm 46\dotarcs8$) within a 5\arcm ($\sim 0.4R_{500}$) radius. We
modeled the diffuse source emission with the thin-plasma emission code
APEC (Smith et al. \cite{smi01}) in XSPEC v12.9.1, leaving
temperature, metal abundance, and normalization as free parameters
(the solar abundances were taken from Asplund et al. \cite{asp09}) and
redshift fixed to the optical value of $z=0.0469$ (see
Sect.~\ref{prop}). This component is absorbed by the Galactic hydrogen
column density along the line of sight, which we fixed to the HI4PI
Map value ($N_{\rm H}= 1.19\times10^{20}$ cm$^{-2}$; Ben Bekhti et
al. \cite{bek16}). We found a best-fit temperature $T_{\rm X}=2.02\pm
0.09$ keV and a metal abundance $Z$=0.26$\pm$0.05 in solar units
(C-Statistic = 1699.93 using 1659 PHA bins and 1653 d.o.f.).

From the scaling relation reported in Arnaud et al. (\cite{arn05},
their Eq.~2) and the computed mean cluster temperature, we estimated
the masses $M_{500} = (1.12\pm 0.15)\times 10^{14}$ \m and $M_{200} =
(1.54\pm 0.25)\times 10^{14}$ \mm, from which we derived $R_{500}=
0.72\pm 0.03$ Mpc $\sim$12.7\arcm and $R_{200}= 1.09\pm 0.07$ Mpc
$\sim$19.2\arcmm. The values of $R_{200}$ and $M_{200}$ estimated from
the X-ray measurements are in agreement within errors with those
derived from the optical properties (see Table~\ref{tabv} and
Table~\ref{tabx}).

\begin{figure}
\centering
\resizebox{\hsize}{!}{\includegraphics{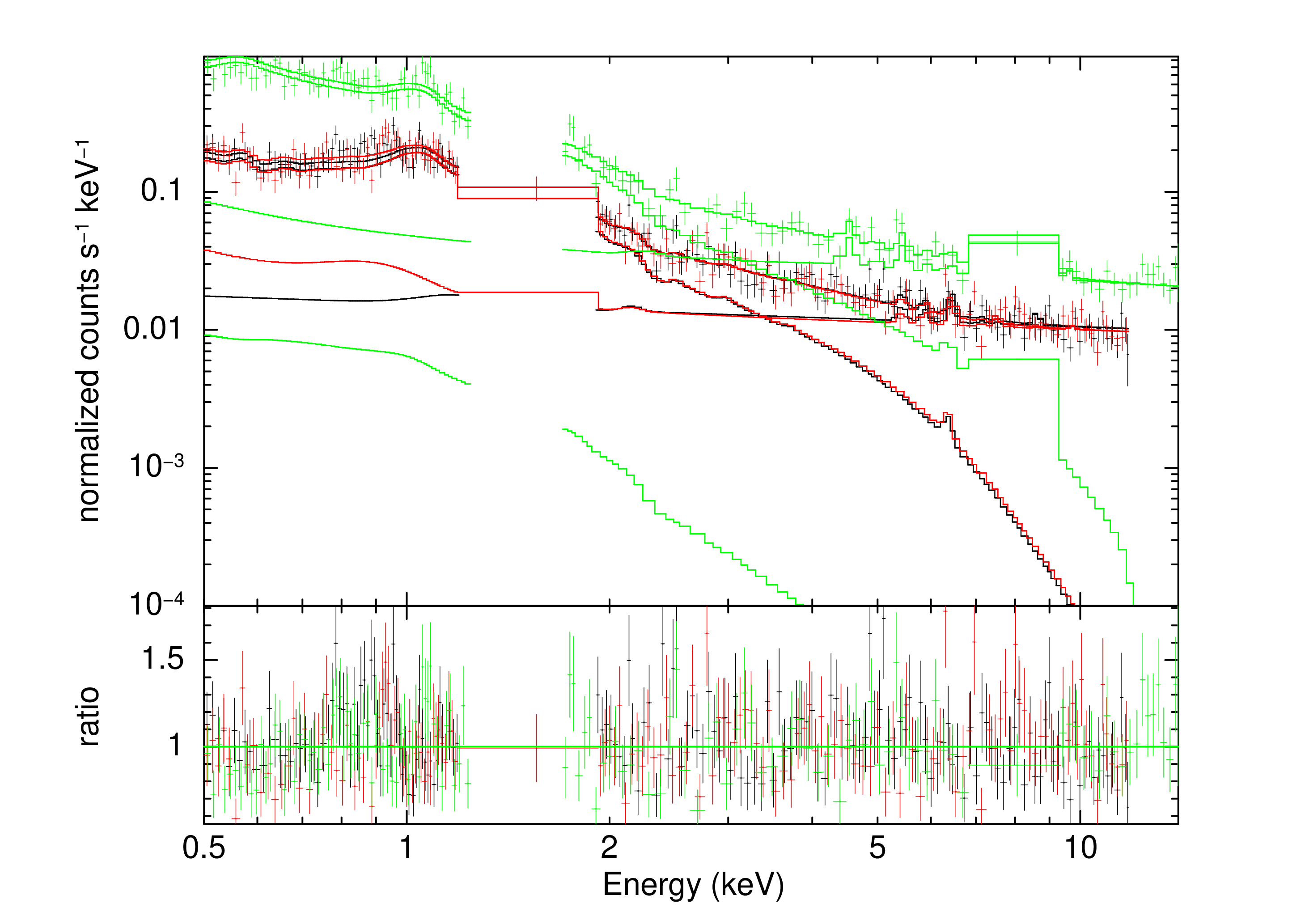}}
\caption{EPIC spectrum of A1213 extracted from the white circle shown
  in Fig.~\ref{figX1}. The circle has a radius of 5\arcmm. Black points
  are MOS1, red points MOS2, and green points pn detector data. The
  lines are the best-fit sky background and source models for the
  three EPIC detectors. The data around 1.5 keV and 7 keV are excluded
  from the analysis to avoid strong instrumental line emissions (see
  Ghirardini et al. \cite{ghi19}).  The bottom panel shows the data
  divided by the folded model.}
\label{figX2}
\end{figure}

\begin{figure}
\centering
\resizebox{\hsize}{!}{\includegraphics{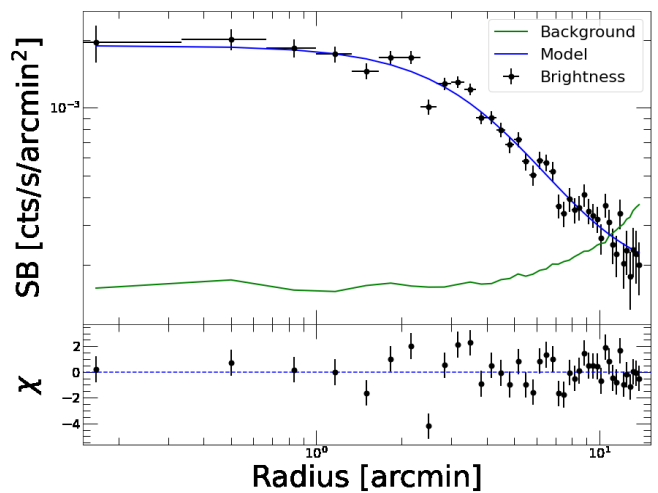}}
\caption{X-ray surface brightness profile of A1213 (black dots). The
  blue curve represents the best fit, using a single $\beta$ model, as
  described in the text. The particle background (shown as the green
  curve) has already been subtracted from the X-ray surface brightness
  profile. The bottom panel shows the contribution of each bin to
  $\chi2$.}
\label{figX4}
\end{figure}

Finally, we extracted the surface brightness (SB) profile of the
cluster up to 14\arcm and then we fitted the profile with a single
$\beta$-model, $I(R)=I_0[1+(R/R_{\rm c})^2]^{-3\beta+0.5}+b$
(Cavaliere \& Fusco-Femiano \cite{cav76}), where $I_0$, $R_{\rm c}$,
$\beta$ and $b$ are the central surface brightness, core radius,
slope, and sky background level, respectively. The extended emission
to the north of the cluster, visible in Fig.~\ref{figX1}, has been
excluded from the extraction of the SB profile. Our best fit gives
$\beta = 0.69\pm 0.20$, $R_{\rm c} = 4.8\pm 1.2$ arcmin, Log$(I_0) =
-2.77\pm 0.03$ and Log$(b) = -3.76\pm 0.13$ ($I_0$ and $b$ in units of
counts s$^{-1}$ arcmin$^{-2}$, $\chi2/{\rm d.o.f.} = 65.8/38$). The SB
profile and the best fit model are plotted in Fig.~\ref{figX4}. This
result allowed us also to reconstruct the X-ray luminosity of the
cluster. In particular, using the {\it pyproffit.deproject} module
(Eckert \cite{eck16} and \cite{eck20}), we computed the luminosity
within $R_{500}$. We obtained $L_{\rm X}\,[0.1-2.4\,{\rm keV}] =
(1.53\pm 0.08)\times 10^{43}$ erg s$^{-1}$.

By using the {\it pyproffit.deproject} module, we also reconstructed
the proton density profile ($n_{\rm p}$) of the ICM. Remembering that
$n_{\rm p}$ is related to the electron density profile by $n_{\rm
  e}=1.17 n_{\rm p}$ (Ghirardini et al. \cite{ghi19}), for the central
electron density, we obtained a value of $n_{\rm
  e,0}=1.19^{+0.43}_{-0.26}\times 10^{-3}$ cm$^{-3}$. Govoni et
al. (\cite{gov17}) found a scattered correlation between $n_{\rm e,0}$
and the mean central magnetic field strength $\left<B_{0}\right>$ (see
their Fig.~15, right panel). This allows us to provide a rough
estimate of the magnetic field in the central region of the
cluster. We find $\left<B_{0}\right>\sim 2-3$ $\mu$G.

\subsection{The group at NNW}
Figure~\ref{figX1} also shows an extended X-ray emission at
$\sim$9\arcm NNW of A1213 and centered on the SDSS galaxy CGCG 156-041
(R.A.=$11^{\mathrm{h}}16^{\mathrm{m}}14\dotsec33$, Dec.=$+29\degree
23\arcmm 06\dotarcs9$). The galaxy is at $z=0.02929$, so it is the
main member of a foreground system, not related to A1213. We extracted
a spectrum from a circular region of 1.5\arcm around this galaxy
(plotted in magenta in Fig.~\ref{figX1}) to estimate the physical
properties of this foreground source. For the source emission, we used
again the APEC model fixing the metal abundance at $Z$=0.25 (solar
units), $z=0.02929$, and $N_{\rm H}=1.22\times 10^{20}$ cm$^{-2}$. The
temperature of the source is $0.83\pm 0.06$ keV, corresponding to a
small group of galaxies with a mass $M_{500}\sim$0.2$\times$10$^{14}$
\m (from the $M_{500}-T_{\rm X}$ relation in Lovisari et
al. \cite{lov21}).

\section{LOFAR radio data}
\label{lofar}
\begin{figure}
\centering
\resizebox{\hsize}{!}{\includegraphics{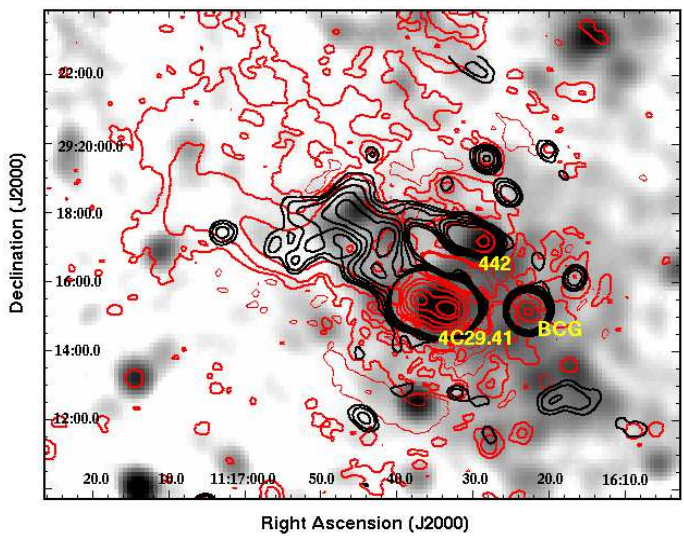}}
\caption
    {LOFAR Radio contours of the diffuse radio emission in A1213 at
      144 MHz (red, HPBW=20\arcss$\times$20\arcss, noise level is 0.15
      mJy beam$^{-1}$, thin contour levels at -0.5 mJy beam$^{-1}$,
      thick contour levels at 0.5, 2, 8, 32, 128, and 1024 mJy
      beam$^{-1}$) and VLA contours at 1.4 GHz (black, as in
      Fig.~\ref{figottico}), superimposed to the smoothed
      \emph{XMM-Newton} image in the energy band 0.7-1.2 keV. Yellow
      labels highlight prominent galaxies mentioned in the text.}
\label{figlofar}
\end{figure}

The field of A1213 is covered by the LOw-Frequency ARray (LOFAR)
Two-metre Sky Survey (LoTSS-DR2; Shimwell et al. \cite{shi22}). These
data are public (Mosaic Field: P168+30) and provide a picture of the
cluster region at 144 MHz, a lower frequency with respect to the VLA
data of G09. As shown first by Hoang et al. (\cite{hoa22}), the LOFAR
images confirm the existence of the diffuse emission observed in NVSS
and VLA data. In particular, at 144 MHz the radio emission extends
toward the east, with a projected size of $\sim$510 kpc, quite longer
than the size of the emission seen by G09 at 1.4 GHz (see
Fig.~\ref{figlofar}). However, we also point out that some fragmented
diffuse emission is detected at the cluster center at the 3.3$\sigma$
level. In particular, there are hints of faint diffuse structures
toward NW and SE of the BCG, to the N of the tailed radio galaxy
(ID~442) and to the SSE of 4C29.41.

The existence of data sets at two different frequencies allowed us to
produce a spectral radio index map. Indeed, spatially resolved
spectral index mapping of diffuse sources can provide useful
information on their nature and origin (e.g., van Weeren et
al. \cite{wee19}).

\begin{figure}
\centering
\resizebox{\hsize}{!}{\includegraphics{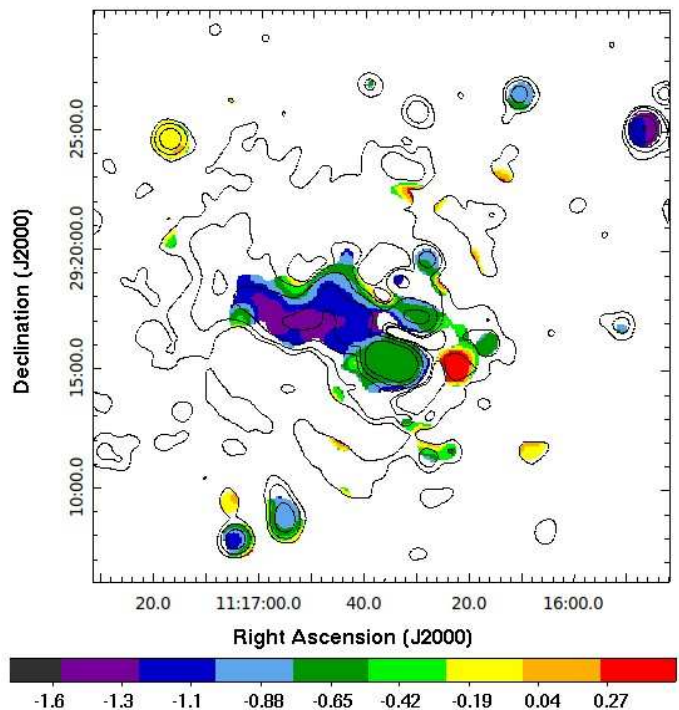}}
\caption{VLA 1.4 GHz/LOFAR 144 MHz spectral index map of the radio
  emission of A1213 at the angular resolution of
  40\arcss$\times$40\arcss. As a reference, we also plot in black the
  contours of the LOFAR image at the same resolution. Contour levels
  are at 0.84, 3.36, 13.44, 53.76, and 215.04 mJy beam$^{-1}$.}
\label{figrsi}
\end{figure}

To produce the spectral index image, we retrieved the LoTTS-DR2 image
at the resolution of 20\arcs from the Mosaic Field P168+30 (see above)
and compared it with the image obtained from the VLA data of G09. In
particular, the VLA image was obtained combining C and C/D
configuration data (see G09 for more details) using the same cell-size
and resolution of the LOFAR image. The two images were put on the same
reference frame using the AIPS\footnote{AIPS is produced and
  maintained by the National Radio Astronomy Observatory, a facility
  of the National Science Foundation operated under cooperative
  agreement by Associated Universities, Inc.} task HGEOM and convolved
to the resolution of 40\arcs to have a better signal-to-noise ratio in
the faint diffuse emission regions. The noise level in the 144 MHz
image is 0.28 mJy beam$^{-1}$ and at 1.4 GHz is 0.25 mJy beam$^{-1}$
in the region of A1213. A clip at the 3$\sigma$ level was applied in
producing the spectral index image to avoid large uncertainties. Our
map (see Fig.~\ref{figrsi}) has small uncertainties on the spectral
index, mainly in the range 2-12\% across the source and indicates a
steepening of the radio spectrum from the north to the south of the
diffuse emission. In particular, a slice in N-S of this region shows
that the spectral index drops from -1.0$\pm$0.04 to
-1.4$\pm$0.04. This steepening is further highlighted by comparing the
average values of the spectral index in the northern and southern
regions of the diffuse emission.

We are aware that the LOFAR data provide a better coverage in the
short spacing range than the VLA data. However, we note that in the
VLA configuration adopted by G09 there are 10 inner antennas in D
configuration and the largest structure that can be imaged at high
sensitivity is 16.2\arcm in size. The diffuse emission detected with
LOFAR is $\sim$13\arcmm$\times$9\arcm in size, significantly smaller
than 16.2\arcmm. Moreover, the size of the structure where we derived
the spectral index distribution is even much smaller. Therefore, we
are confident that the spectral comparison is not affected by the
different uv-coverage (for a similar case see, e.g., Feretti et
al. \cite{fer04}).

The steepening of the radio spectrum in the N-S direction suggests
that this source, rather than a radio halo, could be a radio relic. In
fact, this spectral behavior has been observed in confirmed radio
relics for which spectral index maps are available (e.g. in the
cluster CIZA J2242.8+53.01; van Weeren et al. \cite{wee10}).

\begin{figure}
\centering
\resizebox{\hsize}{!}{\includegraphics{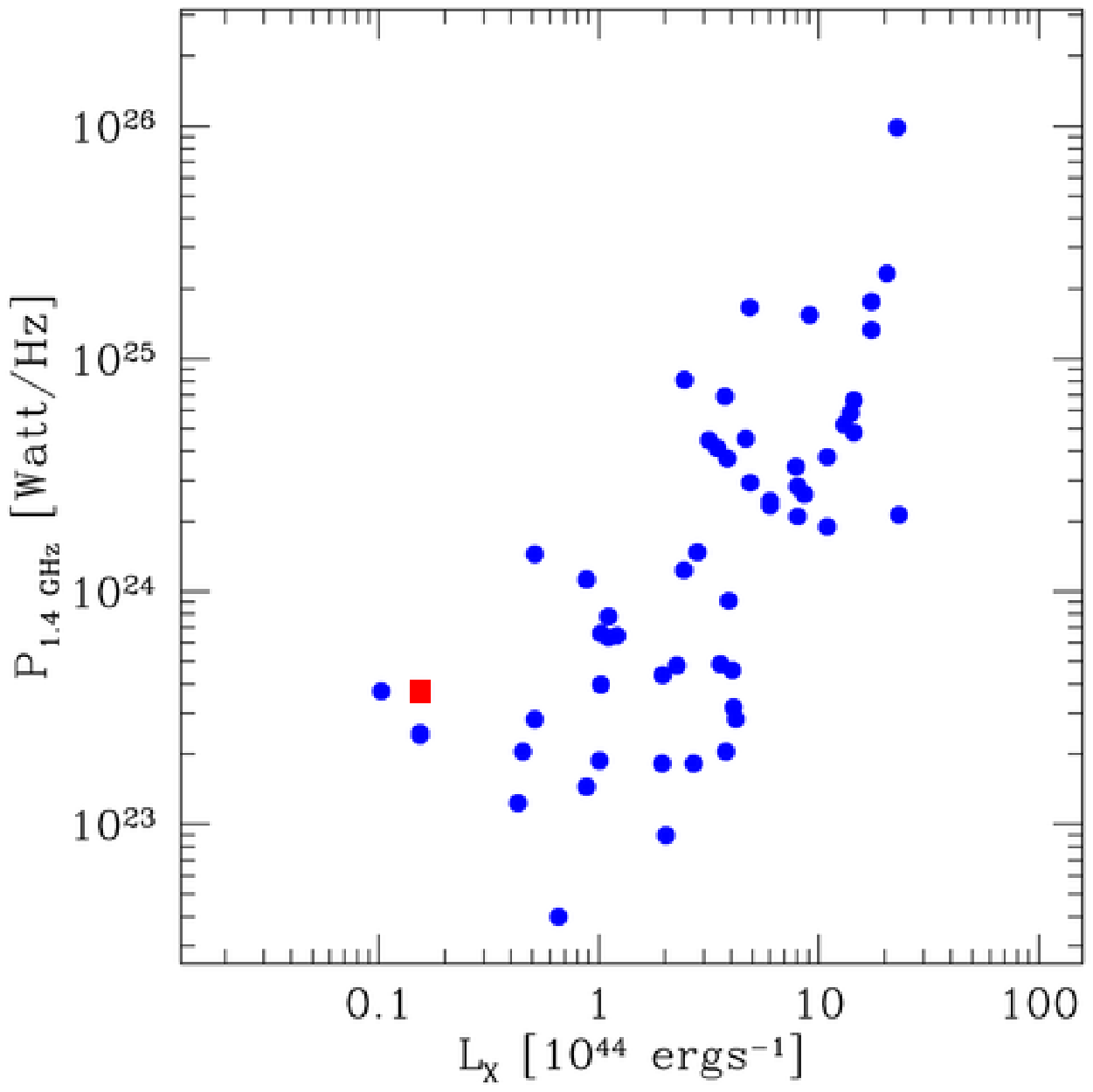}}
\caption{Monochromatic radio power of relics at 1.4 GHz versus the
  cluster X-ray luminosity of the parent clusters (energy band 0.1–2.4
  keV). Blue dots are radio relics from the literature (see text),
  while the red square refers to A1213.}
\label{figprelvslx}
\end{figure}

It is known that the properties of relics are related to the X-ray
properties of their parent clusters (e.g., Feretti et
al. \cite{fer12}). This fact offers a way to compare the suspected
relic of A1213 with the radio relics known in the literature. The
comparison is shown in Fig.~\ref{figprelvslx}, which reports the power
of radio relics at 1.4 GHz versus the X-ray luminosity of the parent
clusters in the energy range 0.1-2.4 keV. Blue dots are radio relics
from the literature (see Feretti et al. \cite{fer12} and van Weeren et
al. \cite{wee19}), while the red square shows the location of A1213 in
the diagram considering the new X-ray luminosity derived in this work.
The figure shows that the diffuse source of A1213 is in agreement with
the relic power - cluster X-ray luminosity correlation.

\section{Summary and discussion}
\label{discu}

The mass estimates based both on the optical and the X-ray data
confirm that A1213 is a poor galaxy cluster ($M_{200}\sim 2\times
10^{14}$\mm). Moreover, we collected convincing evidence that this
cluster is far from being dynamically relaxed.

From the optical point of view, despite the fact that the velocity
distribution of member galaxies does not deviate significantly from
the Gaussianity, a compelling argument in favor of a disturbed
dynamics in A1213 comes from the very significant peculiar velocity of
the BCG. This is quite unusual in regular clusters, where the
brightest galaxy is located at the peak of the velocity distribution
(see, e.g., the case of CL1821+643; Boschin et al. \cite{bos18}).

With regard to the 2D analysis of the galaxy distribution, we can see
that the cluster is only slightly elliptical and roughly oriented in
the NE-SW direction. Our analysis of optical substructures suggests
that A1213 is composed by several galaxy groups. In particular, we
detected two external (at $R\sim R_{200}$) groups: 2D--NW and 2D--NE
(see Table~\ref{tabdedica2dz}). 2D--NW is the densest one and exhibits
a dominant galaxy (ID~274) and several galaxies all around, thus maybe
it has not yet crossed the central region of A1213. On the contrary,
2D--NE seems elongated in the same direction as the X-ray emission
(see below). It could be a group in a recent, post-merger phase
occurring in the plane of the sky.

About the central cluster region, both the velocity gradient and the
DSV test suggest the existence of two substructures oriented in the
NE-SW direction. The one with lower radial velocity is detected also
by the 3D-DEDICA test (3D-CORE in Table~\ref{tabdedica3d}). Indeed,
the core of A1213 is quite intricate. The existence of the BCG and the
bright couple BG1+BG2, which differ substantially in terms of radial
velocity ($\gtrsim$1000 \kss), could suggest that these galaxies trace
a merger almost along the line of sight. In fact, during a merger of
two groups, the external galaxies are usually swept away and only the
dense cores survive (see simulations by, e.g., Gonz\'alez-Casado et
al. \cite{gon94}; Vijayaraghavan et al. \cite{vij15}). However, if the
groups did not have compact cores but rather their luminosity was
concentrated in a bright dominant galaxy (or a couple of bright
galaxies), the merger is not visible as a double peak in the velocity
distribution and it is traced only by the brightest galaxies.

Another important piece of evidence is the fact that blue,
star-forming galaxies (e.g., galaxy ID~441 in
Table~\ref{catalogA1213}) are not restricted to the peripheral regions
of the cluster, as shown from the projected phase space of cluster
galaxies (Fig.~\ref{figvd}). This is not what is commonly found in
galaxy clusters, where blue galaxies usually tend to avoid the core
regions (e.g., Girardi et al. \cite{gir15}; Mercurio et
al. \cite{mer21}) and is consistent with the scenario of an assembling
cluster through accretion of several poor groups rich in late-type
galaxies. Another explanation could be that in A1213, which is a poor
cluster, the ram-pressure by the ions of the ICM (see Boselli et
al. \cite{bos22} for a recent review) is not efficient to quench the
star formation in blue galaxies falling into the cluster core.

As for the X-ray properties of A1213, it displays a patchy, shallow
X-ray emission on scales of $\sim$700 \kpc (see Fig.~\ref{figX1}) in
the presence of various point sources. A comparison with
Fig.~\ref{figottico} shows that several of these sources coincide with
the most prominent optical galaxies in the field, such as BCG, BG1,
and ID~442. None of these bright galaxies coincide with the centroid
of the X-ray emission. Once the point sources detected in the image
are excised, the diffuse X-ray emission appears almost regular,
slightly elongated in the NE$-$SW direction, with an ellipticity of
$\epsilon\sim 0.1$ and $PA\sim 35$\degree within 0.5$R_{500}$.

To describe the morphological state of A1213 in the X-ray band, two
parameters are useful: the surface brightness concentration $c_{\rm
  SB}$ and the centroid shift $w$ (see Santos et al. \cite{san08};
Poole et al. \cite{poo06}; Maughan et al. \cite{mau08} for
details). Since a central surface brightness excess is a primary
indicator of the presence of a cool core (Fabian et al. \cite{fab84}),
then the $c_{\rm SB}$ parameter is a useful indicator of the dynamical
state of the cluster. On the other hand, the $w$ parameter is
sensitive to the presence of bright substructures. For A1213, we
derive $c_{\rm SB}=0.14$ and $w = 0.025$, a clear indication of this
cluster to be unrelaxed. In fact, typical values for disturbed
clusters are $c_{\rm SB}<0.19$ and $w>0.01$, as found by Campitiello
et al. (\cite{cam22}).

In summary, the diffuse X-ray emission overlaps with the optical
galaxy distribution and is slightly elongated in the same NE-SW
direction.  Moreover, the X-ray morphology, the absence of
well-defined emission peaks and the above-defined dynamical indicators
strengthen the evidence that A1213 is in an unrelaxed dynamical
state. We also note that although the diffuse X-ray emission within
$R_{500}$ is clumpy (Fig.~\ref{figX1}), its large-scale shape appears
only slightly elongated. This indicates that if the cluster core had a
recent interaction it probably did not occur on the plane of the sky.

About the X-ray luminosity of A1213, it was previously proposed that
the cluster is underluminous to explain the discrepancy with the power
of the radio halo (G09, Giovannini et al. \cite{gio11}). Considering
our new estimate of the luminosity within $R_{500}$, together with the
value of the gas temperature derived through the spectral analysis
(see Sect.~\ref{xray1}), we can now claim that A1213 is not an
underluminous cluster (e.g., Lovisari et al. \cite{lov21}, their
Fig.~3). Instead, we determine that it is quite a typical poor
cluster.

The purpose of our work was to characterize the optical/X-ray/radio
properties of A1213 in order to study the environment of the diffuse
radio emission discovered by G09 in this cluster and shed new light on
its nature. The SDSS and \emph{XMM-Newton} data show a good agreement
in the cluster central region, including the fact that both data sets
indicate an elongation in the NE-SW direction. Instead, the observed
radio emission does not coincide with the diffuse X-ray emission (see
Fig.~\ref{figottico}), as occurs generally in unrelaxed clusters with
radio halos (e.g., Govoni et al. \cite{gov01}), because of its offset
with respect to the ICM distribution and its extension toward the NE,
the same direction traced by the galaxy distribution from the core to
the cluster periphery (at $R\sim R_{200}$).

Indeed, recent LOFAR images at 144 MHz of the LoTSS-DR2 confirmed the
presence of the diffuse emission observed in A1213 at higher
frequencies (see Hoang et al. \cite{hoa22}, their Fig.~4, as well as
our Fig.~\ref{figlofar}). However, since the radio emission does not
follow the X-ray emission, Hoang et al. (\cite{hoa22}) infer that this
extended source is not a radio halo, but it is the tail of the central
radio galaxy 4C29.41 (our ID~467=BG1) bent by the interaction with the
ICM (see their Fig.~4). This would explain why A1213 is an outlier in
the normal scale relations between X-ray and radio properties of
radio-halo clusters (e.g., Cassano et al. \cite{cas13}). But the
scenario could be even more complicated. In fact, Hoang et
al. (\cite{hoa22}) detect an excess of diffuse emission on the
easternmost region of the emission seen with LOFAR (see their Fig.~4,
left panel). Since there is no obvious optical counterpart of this
source, its origin is unclear and these authors propose that it could
be associated to a merger occurring in the NE-SW direction.

Our VLA 1.4 GHz/LOFAR 144 MHz spectral index map rejects the
hypothesis of the radio galaxy tail and supports a different
explanation. In fact, the spectral index distribution
(Sect.~\ref{lofar} and Fig.~\ref{figrsi}) is compatible with a radio
relic interpretation, where ``fossil'' electrons of the radio galaxy
4C29.41 (but also of the head-tail galaxy ID~442) are reaccelerated by
shock(s) due to a merger. The radio relic is elongated from the
cluster center toward the cluster periphery, therefore it is located
either in front or behind the cluster. This is consistent with a
merger in the N-S or NE-SW directions, in agreement with the results
of our optical analysis. A relic with a similar structure is detected
in the cluster Abell 115 (Govoni et al. \cite{gov01}, Botteon et
al. \cite{bot16}), although the latter is much larger in size. The
relic hypothesis is also supported by the plot shown in
Fig.~\ref{figprelvslx}, where the diffuse source of A1213 fits the
empirical correlation between the power of radio relics and the X-ray
luminosity of the parent clusters.

Finally, Fig.~\ref{figlofar} shows some evidence of fragmented diffuse
radio emissions at the cluster center whose nature is uncertain. They
could be related to the relic, or could be the tip of the iceberg of
very low-surface-brightness emission permeating the cluster center.
Indeed, our estimate of the central magnetic field strength,
$\left<B_{0}\right>\sim 2-3$ $\mu$G (Sect.~\ref{xray1}), is compatible
with the possible presence of a faint radio halo in the cluster core
(e.g., van Weeren et al. \cite{wee19}). Also, recent works (e.g.,
Hoang et al. \cite{hoa21}, Botteon et al. \cite{bot21}) reported the
discovery with LOFAR of radio halos in low-mass ($M_{500}\lesssim
5\times 10^{14}$\mm) galaxy clusters. Thus, the existence of a faint
halo in A1213 ($M_{500}\sim 1\times 10^{14}$\mm), while quite
uncommon, would not be so extraordinary.

In conclusion, A1213 represents an interesting target to investigate
the connection between optical (galaxy population) and X-ray (ICM)
cluster properties and diffuse radio emissions in a low-mass regime
that is still mostly unexplored. In this context, new deeper X-ray
observations of this cluster could be decisive to test the proposed
relic scenario. In fact, the eventual detection of X-ray surface
brightness discontinuities (associated with shocks in the ICM) in
correspondence to the suspected radio relic would strengthen our
interpretation. Instead, dedicated LOFAR LBA observations (e.g., de
Gasperin et al. \cite{deg21}) could be crucial in characterizing the
very low-surface-brightness diffuse emission detected in the center of
the cluster.

\begin{acknowledgements}
We thank the anonymous referee for his/her useful comments and
suggestions.\\
VV acknowledges support from Istituto Nazionale di Astrofisica (INAF)
mainstream project ``Galaxy Cluster Science with LOFAR''
1.05.01.86.05.\\
This research has made use of the galaxy catalog of the Sloan Digital
Sky Survey (SDSS). Funding for the SDSS has been provided by the
Alfred P. Sloan Foundation, the U.S. Department of Energy Office of
Science, and the Participating Institutions. SDSS acknowledges support
and resources from the Center for High-Performance Computing at the
University of Utah. The SDSS web site is http://www.sdss.org/, where
the complete list of the funding organizations and collaborating
institutions can be found.\\
This paper is also based on archival observations obtained with
\emph{XMM-Newton}, an ESA science mission with instruments and contributions
directly funded by ESA Member States and the USA (NASA).\\
This paper is based in part on data obtained with the International
LOFAR Telescope (ILT) under project code LT10$_{-}$010. LOFAR (van
Haarlem et al. 2013) is the LOw Frequency ARray designed and
constructed by ASTRON. It has observing, data processing, and data
storage facilities in several countries, which are owned by various
parties (each with their own funding sources), and which are
collectively operated by the ILT foundation under a joint scientific
policy. The ILT resources have benefited from the following recent
major funding sources: CNRS-INSU, Observatoire de Paris and Université
d'Orléans, France; BMBF, MIWF-NRW, MPG, Germany; Science Foundation
Ireland (SFI), Department of Business, Enterprise and Innovation
(DBEI), Ireland; NWO, The Netherlands; The Science and Technology
Facilities Council, UK; Ministry of Science and Higher Education,
Poland; The Istituto Nazionale di Astrofisica (INAF), Italy.\\
This research made use of the Dutch national e-infrastructure with
support of the SURF Cooperative (e-infra 180169) and the LOFAR e-infra
group. The Jülich LOFAR Long Term Archive and the German LOFAR network
are both coordinated and operated by the Jülich Supercomputing Centre
(JSC), and computing resources on the supercomputer JUWELS at JSC were
provided by the Gauss Centre for Supercomputing e.V. (grant CHTB00)
through the John von Neumann Institute for Computing (NIC).\\
This research made use of the University of Hertfordshire
high-performance computing facility and the LOFAR-UK computing
facility located at the University of Hertfordshire and supported by
STFC [ST/P000096/1], and of the Italian LOFAR IT computing
infrastructure supported and operated by INAF, and by the Physics
Department of Turin university (under an agreement with Consorzio
Interuniversitario per la Fisica Spaziale) at the C3S Supercomputing
Centre, Italy.

\end{acknowledgements}

\begin{table*}
  \caption[]{Prominent member galaxies of A1213. Table lists
    galaxies with magnitude $r\leq r_{\rm BCG}+1$.}
        \label{catalogA1213}
              $$ 
            \begin{array}{r c c r r l}
            \hline
            \noalign{\smallskip}
            \hline
            \noalign{\smallskip}

\mathrm{ID} & \alpha,\delta\,(\mathrm{J}2000) & r\, &V\,& \Delta V& \mathrm{Comments}\\
 &  \mathrm{h:m:s,\degree:\arcmm:\arcs} & &\mathrm{\,\,(\,km}&\mathrm{s^{-1}\,)\,\,}& \\
            \hline
            \noalign{\smallskip}
422        & 11\ 16\ 22.70,+29\ 15\ 08.3&13.64&  13575&4&{\rm BCG,\,radio galaxy}\\
467        & 11\ 16\ 34.61,+29\ 15\ 17.2&14.41&  14593&3&{\rm BG1,\,FRII\,radio\,galaxy\,4C29.41}\\
468        & 11\ 16\ 35.03,+29\ 14\ 58.8&14.61&  14988&3&{\rm BG2}\\
441        & 11\ 16\ 28.07,+29\ 19\ 36.1&13.96&  13863&2&{\rm Spiral\,radio\,galaxy}\\
274        & 11\ 15\ 10.14,+29\ 31\ 48.2&14.11&  14871&3&{\rm Far\,from\,the\,center;\,in\,substructure\,2D-NW}\\
408        & 11\ 16\ 17.41,+29\ 13\ 35.1&14.21&  13112&3&{\rm }\\
531        & 11\ 17\ 00.39,+29\ 08\ 26.2&14.40&  13977&4&{\rm }\\
 87        & 11\ 13\ 19.44,+29\ 48\ 39.9&14.41&  13542&3&{\rm In\,substructure\,2D-NW}\\
215        & 11\ 14\ 34.06,+29\ 32\ 42.5&14.51&  14031&3&{\rm In\,substructure\,2D-NW}\\
442        & 11\ 16\ 28.46,+29\ 17\ 09.4&14.64&  14191&4&{\rm Head-tail\,radio\,galaxy}\\   
\noalign{\smallskip}                                  
            \hline                                          
            \end{array}                                  
            $$ \tablefoot{Data from SDSS DR13. Listed: identification
              number of each galaxy, ID; right ascension and
              declination, $\alpha$ and $\delta$ (J2000); $r$
              magnitude; heliocentric radial velocities, $V=cz$, with
              errors, $\Delta V$; comments.}

\end{table*}

\begin{table*}
        \caption{Global properties of A1213 as inferred from the analysis of the optical data.}
         \label{tabv}
            $$
         \begin{array}{l r}
            \hline
            \noalign{\smallskip}
            \hline
            \noalign{\smallskip}

            N_{\rm gal} & 143\\
            \noalign{\smallskip}
            ^{\mathrm{a}}\alpha,\delta\ (\mathrm{J}2000)& 11{\rm:}16{\rm:}22.70,\ +29{\rm:}15{\rm:}08.3\\
            \noalign{\smallskip}
            \mathrm{\left<V\right>}\ ({\rm km}\ {\rm s}^{-1})& 14052\pm39\\
            \noalign{\smallskip}
            \sigma_{\rm V}\ ({\rm km}\ {\rm s}^{-1})& 463_{-31}^{+41}\\
            \noalign{\smallskip}
            N_{\rm gal}\ (<R_{200})& 81\\
            \noalign{\smallskip}
            \sigma_{\rm V,200}\ ({\rm km}\ {\rm s}^{-1})& 573_{-38}^{+52}\\
            \noalign{\smallskip}
            R_{200}\ ({\rm Mpc})& 1.20_{-0.08}^{+0.11}\\
            \noalign{\smallskip}
            M_{200}\ (10^{14}\ {\rm M}_{\odot})& 2.0_{-0.4}^{+0.6}\\
         \noalign{\smallskip}

              \noalign{\smallskip}
            \hline
            \noalign{\smallskip}
         \end{array}\\         
         $$
         
\begin{list}{}{}
\item[$^{\mathrm{a}}$] As an optical center, we list the position of the BCG. 
\end{list}
\end{table*}

\begin{table}
        \caption{Optical substructures detected with 2D-DEDICA.}
         \label{tabdedica2dz}
            $$
         \begin{array}{l r c c c}
            \hline
            \noalign{\smallskip}
            \hline
            \noalign{\smallskip}
\mathrm{Group} & N_{\rm gal} & \alpha,\delta\,(\mathrm{J}2000)&\rho&\mathrm{\left<V\right>}\\
& & \mathrm{h:m:s,\degree:\arcmm:\arcs}&&\mathrm{km\ s^{-1}}\\
         \hline
         \noalign{\smallskip}
\mathrm{2D-CORE} & 56&11\ 16\ 29.2+29\ 15\ 26&1.00&14045\pm\phantom{1}88 \\ 
\mathrm{2D-NW}   & 23&11\ 15\ 22.8+29\ 30\ 55&0.22&13981\pm\phantom{1}61\\            
\mathrm{2D-NE}   & 20&11\ 17\ 33.8+29\ 27\ 46&0.17&14129\pm112 \\           
\hline
              \noalign{\smallskip}
              \noalign{\smallskip}
            \noalign{\smallskip}
         \end{array}
$$

\end{table}

\begin{table}
        \caption{Optical substructures detected with 3D-DEDICA.}
         \label{tabdedica3d}
            $$
         \begin{array}{l r r c c }
            \hline
            \noalign{\smallskip}
            \hline
            \noalign{\smallskip}
\mathrm{Group} & N_{\rm gal} & V_{\rm peak}& \alpha,\delta\,(\mathrm{J}2000)&\rho\\
& & {\rm km\,s^{-1}}&\mathrm{h:m:s,\degree:\arcmm:\arcs}&\\
         \hline
         \noalign{\smallskip}
\mathrm{3D-NW}   &107&13985&11\ 16\ 20.2+28\ 17\ 31&1.00\\
\mathrm{3D-SE}   & 26&13868&11\ 18\ 15.7+29\ 51\ 43&0.25\\
\mathrm{3D-CORE} & 10&13158&11\ 16\ 21.9+29\ 12\ 43&0.23\\
\hline
              \noalign{\smallskip}
              \noalign{\smallskip}
            \noalign{\smallskip}
         \end{array}
$$

\end{table}

\begin{table*}
        \caption{Global properties of A1213 as inferred from the analysis of the X-ray data.}
         \label{tabx}
            $$
         \begin{array}{l r}
            \hline
            \noalign{\smallskip}
            \hline
            \noalign{\smallskip}

            T_{\mathrm{X}}\ ({\rm keV}) & 2.02\pm 0.09\\
            \noalign{\smallskip}
            ^{\mathrm{a}}\alpha,\delta\,(\mathrm{J}2000)& 11{\rm:}16{\rm:}26.40,\ +29{\rm:}14{\rm:}46.8\\
            \noalign{\smallskip}
            Z\ ({\rm solar\ units})& 0.26\pm 0.05\\
            \noalign{\smallskip}
            R_{500}\ ({\rm Mpc})& 0.72\pm 0.03\\
            \noalign{\smallskip}
            M_{500}\ (10^{14}\ {\rm M}_{\odot})& 1.12\pm 0.15\\
            \noalign{\smallskip}
            R_{200}\ ({\rm Mpc})& 1.09\pm 0.07\\
            \noalign{\smallskip}
            M_{200}\ (10^{14}\ {\rm M}_{\odot})& 1.54\pm 0.25\\
            \noalign{\smallskip}
            L_{\rm X}\ [0.1-2.4\,{\rm keV}]\ (<R_{500},\ 10^{43}\ {\rm
              erg}\ {\rm s}^{-1})& 1.53\pm 0.08\\
         \noalign{\smallskip}

              \noalign{\smallskip}
            \hline
            \noalign{\smallskip}
         \end{array}\\         
         $$
\begin{list}{}{}
\item[$^{\mathrm{a}}$] Centroid of the cluster X-ray emission. 
\end{list}

\end{table*}

\end{document}